\documentclass[11pt]{article}

\usepackage[preprint]{acl}

\usepackage{times}
\usepackage{latexsym}

\usepackage[T1]{fontenc}

\usepackage[utf8]{inputenc}

\usepackage{microtype}

\usepackage{inconsolata}

\usepackage{graphicx}

\usepackage{booktabs}
\usepackage{multirow}

\usepackage{amsmath}
\usepackage{amssymb}

\usepackage[normalem]{ulem}

\usepackage{xcolor}

\usepackage{listings}
\title{Omni-Embed-Audio: Leveraging Multimodal LLMs for Robust Audio-Text Retrieval}

\author{HaeJun Yoo, Yongseop Shin, Insung Lee, Myoung-Wan Koo, and Du-Seong Chang \\
  Sogang University \\
  \texttt{\{judejiwoo, sysky309, dlstjd6474, mwkoo, dschang\}@sogang.ac.kr} \\}

\begin{document}
\maketitle

\begin{abstract}
Audio-text retrieval systems based on Contrastive Language-Audio Pretraining (CLAP) achieve strong performance on traditional benchmarks; however, these benchmarks rely on caption-style queries that differ substantially from real-world search behavior, limiting their assessment of practical retrieval robustness. We present Omni-Embed-Audio (OEA), a retrieval-oriented encoder leveraging multimodal LLMs with native audio understanding. To systematically evaluate robustness beyond caption-style queries, we introduce User-Intent Queries (UIQs)---five formulations reflecting natural search behaviors: questions, commands, keyword tags, paraphrases, and exclusion-based negative queries. For negative queries, we develop a hard negative mining pipeline and propose discrimination metrics (HNSR, TFR) assessing models' ability to suppress acoustically similar distractors. Experiments on AudioCaps, Clotho, and MECAT show that OEA achieves comparable text-to-audio retrieval performance to state-of-the-art M2D-CLAP, while demonstrating clear advantages in two critical areas: (1) \textbf{dominant text-to-text retrieval} (+22\% relative improvement), and (2) \textbf{substantially superior hard negative discrimination} (+4.3\%p HNSR@10, +34.7\% relative TFR@10). Mechanism ablations attribute these discrimination gains chiefly to OEA's audio embeddings, which place confusable clips significantly farther apart; the exclusion signal itself rides on query word order---an order sensitivity that cross-model controls show OEA shares with CLAP text encoders rather than uniquely possessing.
\end{abstract}


\section{Introduction}

\begin{figure}[t]
\centering
\includegraphics[width=\columnwidth]{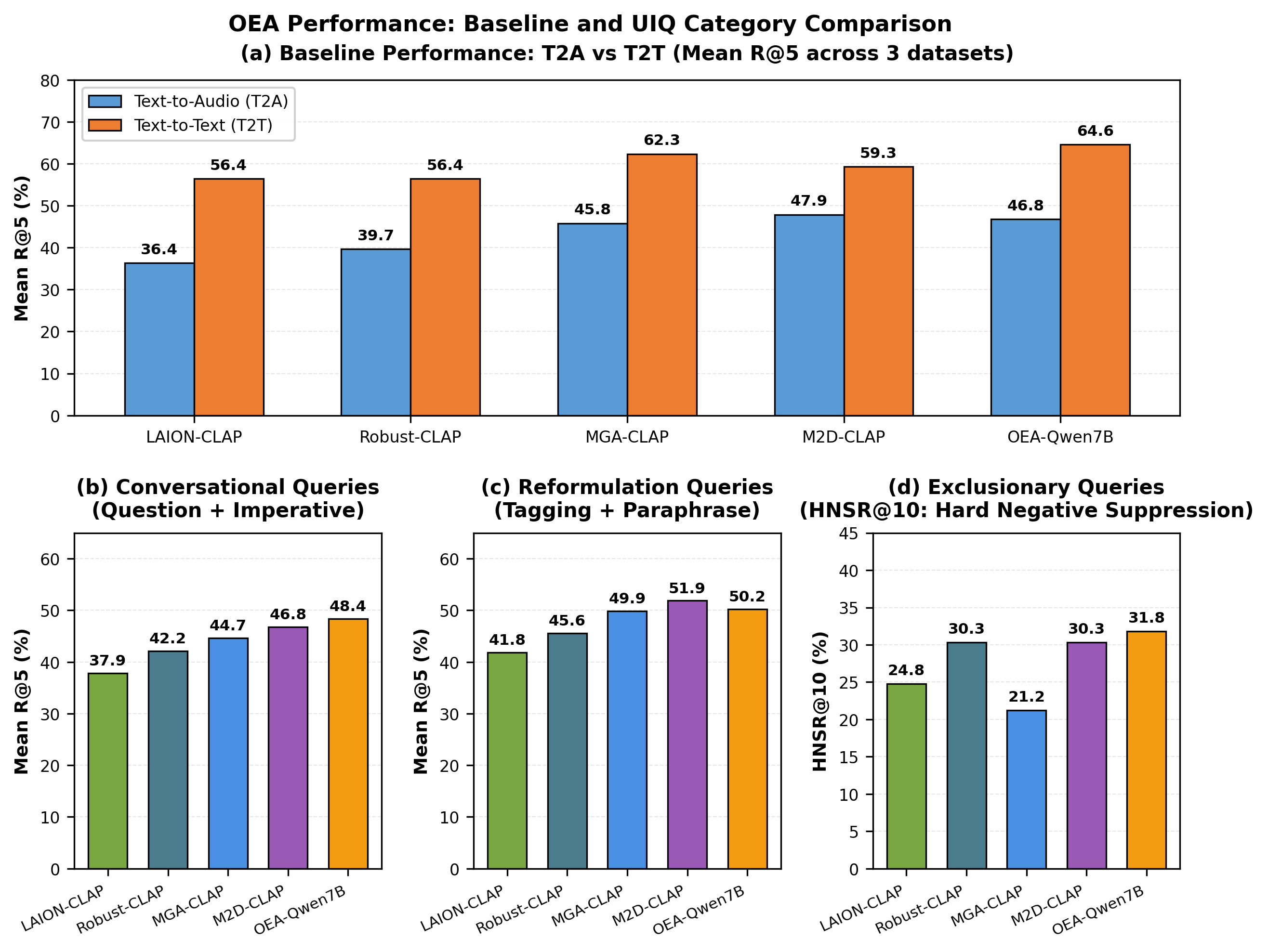}
\caption{Performance comparison using OEA-Qwen7B (+Cl) as representative model (mean R@5 across AudioCaps, Clotho, and MECAT). (a) Baseline performance: While M2D-CLAP leads T2A (47.9\%), OEA achieves competitive results (46.4\%) and substantially outperforms all baselines on T2T (64.8\% vs. M2D-CLAP 59.3\%, +5.5). (b--d) UIQ analysis: M2D-CLAP shows strong UIQ retrieval; however, OEA achieves best Imperative query performance (49.9\% vs. 44.7\%) and substantially outperforms on hard negative discrimination metrics (HNSR@10: 34.6\% vs. 30.3\%, +4.3), demonstrating superior semantic understanding of complex queries.}
\label{fig:main_comparison}
\end{figure}

\begin{figure*}[t]
\centering
\includegraphics[width=0.8\textwidth]{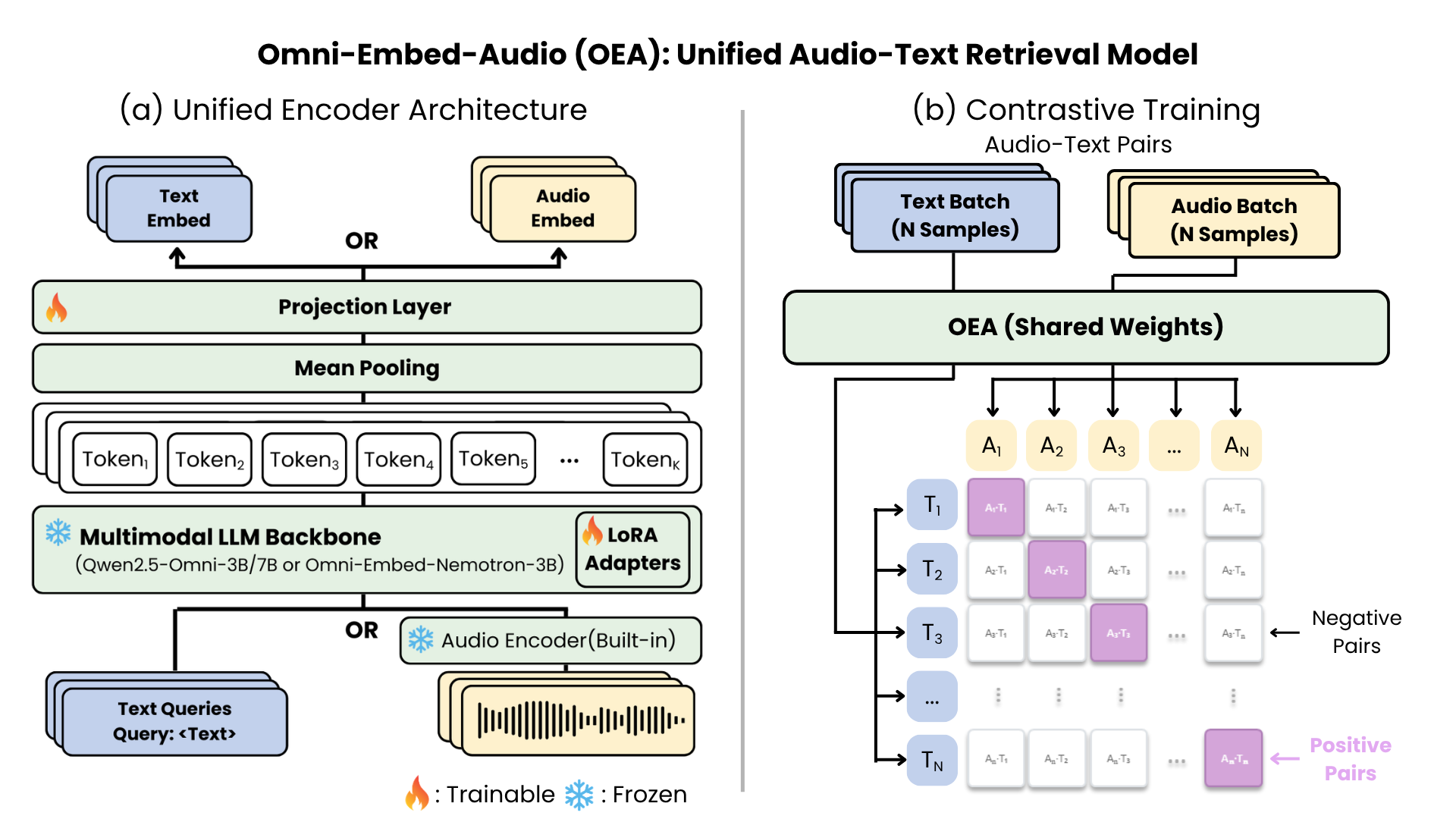}
\caption{OEA architecture overview. A shared multimodal LLM backbone processes both text and audio inputs. LoRA adapters are applied to attention layers, and modality-specific projection heads map representations to a shared 512-dimensional embedding space with L2 normalization.}
\label{fig:oea_architecture}
\end{figure*}

Text-based audio retrieval has emerged as a critical capability for navigating the exponentially growing repositories of audio content across multimedia production, gaming, filmmaking, and creative industries \cite{font2013freesound, koepke2023audio}. Open-vocabulary audio-language models (ALMs), particularly those based on Contrastive Language-Audio Pretraining (CLAP) \cite{niizumi2025m2dclap, li2024advancing, wu2023large, elizalde2023clap}, have achieved remarkable success by learning joint embedding spaces that align audio and text representations. These models enable users to retrieve relevant audio using natural language queries, offering an intuitive interface for searching through vast audio databases.

Despite significant progress on standard benchmarks, recent studies have revealed fundamental limitations in how audio retrieval systems are evaluated. Standard benchmarks such as AudioCaps and Clotho rely on caption-style queries that closely mirror training data distributions, creating an evaluation paradigm that may not reflect real-world retrieval performance. \citet{selvakumar2024robust} demonstrated that existing ALMs suffer performance degradations of up to 16\% when faced with paraphrased queries, while \citet{weck2024language} found substantial mismatch between real-world query patterns (averaging 1.8 tokens) and the descriptive captions used for training. These findings expose a critical gap: current benchmarks may overestimate model capabilities because they test on caption-style queries rather than the diverse \textit{query-type variations} characterizing real-world usage.

To address these limitations, we present \textbf{Omni-Embed-Audio (OEA)}, a retrieval-oriented multimodal encoder that leverages pre-trained LLMs with native audio understanding. Unlike traditional dual-encoder approaches, OEA employs a single shared transformer backbone for both modalities, combined with parameter-efficient LoRA adaptation. We hypothesize that this LLM-based approach can benefit from superior language understanding capabilities, potentially yielding improved performance even on traditional text-to-audio retrieval benchmarks.

To systematically evaluate retrieval robustness, we introduce \textbf{User-Intent Queries (UIQs)}---five query formulations organized into three categories reflecting distinct aspects of real-world search behavior: (1) \textit{Conversational Queries} (questions, commands) anticipating the shift toward speech-based AI interfaces; (2) \textit{Reformulation Queries} (tags, paraphrases) testing robustness to query variation; and (3) \textit{Exclusionary Queries} (negative queries) assessing fine-grained semantic discrimination.

Extensive experiments demonstrate that while recent CLAP models like M2D-CLAP achieve strong text-to-audio retrieval performance, OEA provides complementary strengths: (1) dominant text-to-text retrieval performance (+5.5\%p over M2D-CLAP), (2) superior performance on imperative queries (+5.1\%p), and critically, (3) substantially improved hard negative discrimination (+4.3\%p HNSR@10). These discrimination improvements emerge without explicit UIQ training; our mechanism ablations (Appendix~\ref{sec:mechanism_ablations}) show they arise chiefly from more separable audio embeddings, combined with an order-sensitive encoding of exclusion structure that cross-model controls reveal to be shared with CLAP text encoders.

Our contributions are:
\begin{enumerate}
    \item We present \textbf{OEA}, a unified encoder architecture leveraging multimodal LLMs, demonstrating dominant T2T performance and superior semantic understanding of complex queries.
    \item We identify \textbf{query-type robustness} as a critical dimension of audio retrieval evaluation and introduce the \textbf{UIQ benchmark} with five query types organized into three linguistically-motivated categories.
    \item We propose \textbf{novel hard negative-based evaluation metrics} (HNSR, TFR, $\Delta$-Rank) that reveal limitations in standard retrieval metrics and demonstrate OEA's superior discrimination capabilities.
\end{enumerate}

To facilitate future research, we release three UIQ benchmark datasets (\textbf{AudioCaps-UIQ}, \textbf{Clotho-UIQ}, and \textbf{MECAT-UIQ}) along with an interactive web demo for exploring model behavior across query types.\footnote{UIQ benchmarks: 13,053 queries total (AudioCaps: 4,530, Clotho: 4,722, MECAT: 3,801). Web demo includes 75 representative samples (15 per query type): \url{https://omni-embed-audio.github.io}}


\section{Related Work}

\subsection{Audio-Text Retrieval}

Contrastive Language-Audio Pretraining (CLAP) has become the dominant paradigm for audio-text retrieval \cite{li2024advancing, wu2023large, elizalde2023clap}. LAION-CLAP \cite{wu2023large} scaled training to 630K audio-text pairs with feature fusion and keyword-to-caption augmentation. MGA-CLAP \cite{li2024advancing} introduced multi-granularity aggregation for improved alignment, while CompA \cite{ghosh2024compa} addressed compositional reasoning through composition-aware hard negatives. Most recently, M2D-CLAP \cite{niizumi2025m2dclap} combines self-supervised masked modeling (M2D) with CLAP, achieving state-of-the-art text-to-audio retrieval performance by learning general-purpose audio-language representations that excel in both zero-shot and transfer learning scenarios. However, these models are trained and evaluated primarily on caption-style queries, leaving their robustness to diverse query formulations unexplored.

\subsection{Real-World Query Patterns and Benchmark Limitations}

Standard benchmarks such as AudioCaps \cite{kim2019audiocaps} and Clotho \cite{drossos2019clotho} rely on caption-style queries that mirror training distributions. However, \citet{weck2024language} found that real Freesound queries average only 1.8 tokens, and \citet{penha2022evaluating} demonstrated $\sim$20\% effectiveness drops with paraphrased queries, revealing that single-format benchmarks mask significant model weaknesses.

\subsection{Query Robustness and Negation in Retrieval}

Query sensitivity is well-documented: \citet{selvakumar2024robust} proposed RobustCLAP achieving 0.8--13\% improvements on paraphrases. For negation, NevIR \cite{weller2023nevir} and ExcluIR \cite{zhang2024excluir} showed that neural retrievers struggle with exclusionary queries, with performance barely above random. These findings motivate our UIQ benchmark and hard negative discrimination metrics.


\section{Methodology}

\subsection{OEA Architecture}

Omni-Embed-Audio (OEA) is a retrieval-oriented multimodal encoder designed for audio-text retrieval. The key insight is to leverage pre-trained large language models (LLMs) with native audio understanding capabilities as a unified encoder for both text and audio modalities.

\paragraph{Unified Encoder Architecture}
Unlike traditional dual-encoder approaches that use separate encoders for text and audio, OEA uses a \textbf{single shared transformer backbone} for both modalities (Figure~\ref{fig:oea_architecture}). Text and audio inputs are processed through the same transformer, reducing the modality gap and enabling audio understanding to benefit from the LLM's rich language priors.

\paragraph{Input Processing}
For text encoding, queries are wrapped with a \texttt{query:} prefix, tokenized, passed through the transformer, and mean-pooled over the last hidden layer. For audio encoding, raw waveforms (16kHz mono) are processed through the model's native audio encoder with a \texttt{passage:} prefix and identical mean pooling, producing modality-agnostic representations.

\paragraph{Parameter-Efficient Adaptation}
We employ LoRA adaptation with lightweight adapter matrices attached to attention layers, along with modality-specific projection heads that compress backbone representations into a shared 512-dimensional retrieval embedding space. All backbone weights remain \textbf{frozen}; only LoRA adapters and projection heads are trained, yielding approximately 11--16M trainable parameters.

\paragraph{Training Objective}
We use symmetric contrastive learning with InfoNCE loss ($\tau = 0.07$). Implementation details are provided in Appendix~\ref{sec:implementation_details}.

\paragraph{Backbone Models}
We instantiate OEA with three backbone scales: Omni-Embed-Nemotron-3B \cite{xu2025omniembed} ($\sim$3B parameters), Qwen2.5-Omni-3B \cite{xu2025qwen25omni} ($\sim$3B parameters), and Qwen2.5-Omni-7B \cite{xu2025qwen25omni} ($\sim$7B parameters). All backbones are multimodal LLMs with native audio understanding. A detailed inference efficiency comparison with CLAP baselines is provided in Appendix~\ref{sec:efficiency_analysis}.


\subsection{User-Intent Queries (UIQs)}

\subsubsection{UIQ Taxonomy}

To address the gap between benchmark queries and real-world search behavior identified in Section 2.2, we introduce \textbf{User-Intent Queries (UIQs)}: five query formulations organized into three categories reflecting distinct aspects of real-world search behavior.

\paragraph{Category 1: Conversational Queries}
These query types anticipate the shift from text-based to speech-based AI interfaces, where users interact through natural dialogue rather than keyword input.

\begin{itemize}
\item \textbf{Question Query}: Natural language questions common in conversational search and voice assistants (e.g., \textit{``Can you find clear dog barks echoing in a large hall?''}).

\item \textbf{Imperative Query}: Direct command-style queries reflecting how users interact with AI assistants (e.g., \textit{``Find crisp footsteps on gravel with light echo''}). LLM backbones pre-trained on instruction-following data should excel at parsing such commands.
\end{itemize}

\paragraph{Category 2: Reformulation Queries}
These query types test robustness to the natural variation in how users express identical information needs, addressing the vocabulary mismatch problem central to information retrieval \cite{carpineto2012survey}.

\begin{itemize}
\item \textbf{Keyphrase Query}: Keyword-style queries with comma-separated tags preferred by users seeking concise searches (e.g., \textit{``dog barks, echoing hall, reverberant''}). This format reflects actual Freesound query patterns and tests whether models rely on syntactic structure.

\item \textbf{Paraphrase Query}: Query reformulation has been shown to significantly improve retrieval performance in information retrieval systems \cite{ma2023queryrewriting, jang2024itercqr}. Motivated by this, we include paraphrase queries---declarative descriptions that rephrase audio content using varied vocabulary and structure (e.g., \textit{``Echoing dog barks resonate through a large empty hall''})---to evaluate whether retrieval models maintain robustness when queries are reformulated while preserving semantic intent.
\end{itemize}

\paragraph{Category 3: Exclusionary Queries}
This category tests fine-grained semantic discrimination through queries specifying both desired and excluded content (see Figure~\ref{fig:exclusionary_metrics} for an illustrative example).

\begin{itemize}
\item \textbf{Negative Query}: Queries specifying desired content AND explicit exclusions (e.g., \textit{``Heavy rain and wind on metal surfaces without thunder or engine noise''}). Each negative query is grounded in a pre-mined (target audio, hard negative audio) pair, enabling quantitative evaluation of exclusion understanding.
\end{itemize}

\begin{table}[t]
\centering
\footnotesize
\begin{tabular}{l|cc|cc}
\toprule
\multirow{2}{*}{\textbf{Query Type}} & \multicolumn{2}{c|}{\textbf{Human}} & \multicolumn{2}{c}{\textbf{LLM}} \\
& Mean & Std & Mean & Std \\
\midrule
Question & 4.26 & 0.91 & 4.73 & 0.44 \\
Imperative & 4.16 & 1.06 & 4.33 & 0.87 \\
Keyphrase & 4.35 & 1.05 & 4.47 & 0.50 \\
Paraphrase & 4.14 & 1.21 & 4.67 & 0.47 \\
Negative & 3.82 & 1.25 & 3.93 & 0.57 \\
\midrule
\textbf{Overall} & \textbf{4.15} & 1.12 & \textbf{4.43} & 0.66 \\
\bottomrule
\end{tabular}
\caption{UIQ benchmark validation comparing human evaluation (9 annotators, 675 ratings) with LLM evaluation (Claude Opus 4.5).}
\label{tab:uiq_validation}
\end{table}

\subsubsection{UIQ Generation Process}

For positive query types (Question, Imperative, Paraphrase, Keyphrase), we generate queries using GPT-5.1 with vocabulary grounding constraints requiring queries to reuse wording from original captions. To avoid length-induced distribution shift in evaluation, we maintain query lengths within $\pm$2 words of original captions, preserving semantic completeness for controlled evaluation. For negative queries, we additionally provide hard negative captions to craft precise exclusion statements. Generated queries were not exhaustively reviewed by humans; instead, benchmark quality is assessed post-hoc through sample-based human and LLM validation (Section~\ref{sec:uiq_validation}).

\subsubsection{UIQ Benchmark Validation}
\label{sec:uiq_validation}

We construct three evaluation datasets: \textbf{AudioCaps-UIQ}, \textbf{Clotho-UIQ}, and \textbf{MECAT-UIQ}, which are publicly released together with our code and checkpoints. We validate these benchmarks through both LLM-based evaluation (Claude Opus 4.5) and human annotation, using the same evaluation criterion: \textit{``Does this make sense as a search query regarding the original captions?''} Responses are rated on a 5-point Likert scale (see Appendix~\ref{sec:uiq_validation_appendix} for scale definitions).

\paragraph{Evaluation Methodology}
We validate UIQ quality through two complementary approaches: (1) \textbf{Human evaluation}: 9 annotators rated 75 UIQ samples (15 audio clips $\times$ 5 UIQ types) on a 5-point Likert scale, yielding 675 total ratings; (2) \textbf{LLM-based evaluation}: Claude Opus 4.5 evaluated the same samples using identical criteria.

\paragraph{Validation Results}
Table~\ref{tab:uiq_validation} presents validation results. Human evaluation shows strong validity across all UIQ types (mean 4.15/5.0), with Question and Keyphrase queries receiving highest ratings. LLM evaluation demonstrates similar patterns with moderate Human-LLM agreement. Imperative queries show strongest agreement ($r=0.89$, $p<0.001$). We note that this validation is sample-based: 75 of the 13{,}053 released queries were rated, so the reported scores are estimates of overall benchmark quality rather than the result of an exhaustive audit. We further analyze potential synthetic bias in UIQ generation in Appendix~\ref{sec:uiq_bias_analysis}.


\subsection{Hard Negative Mining and Evaluation Metrics}
\label{sec:hard_negative_mining}

Evaluating exclusionary query understanding requires paired examples where models must distinguish acoustically similar but semantically distinct audio. We develop a four-stage hard negative mining pipeline that combines acoustic similarity retrieval (using MGA-CLAP embeddings) with semantic dissimilarity filtering (using BGE sentence embeddings), yielding audio pairs that are acoustically confusable but semantically distinct. To mitigate single-model dependency, human listening on sampled pairs was used to calibrate the mining thresholds so that retained pairs are perceptually confusable (Appendix~\ref{sec:hn_human_validation}). Details are provided in Appendix~\ref{sec:hn_mining_appendix}.

\begin{figure}[t]
\centering
\includegraphics[width=0.5\textwidth]{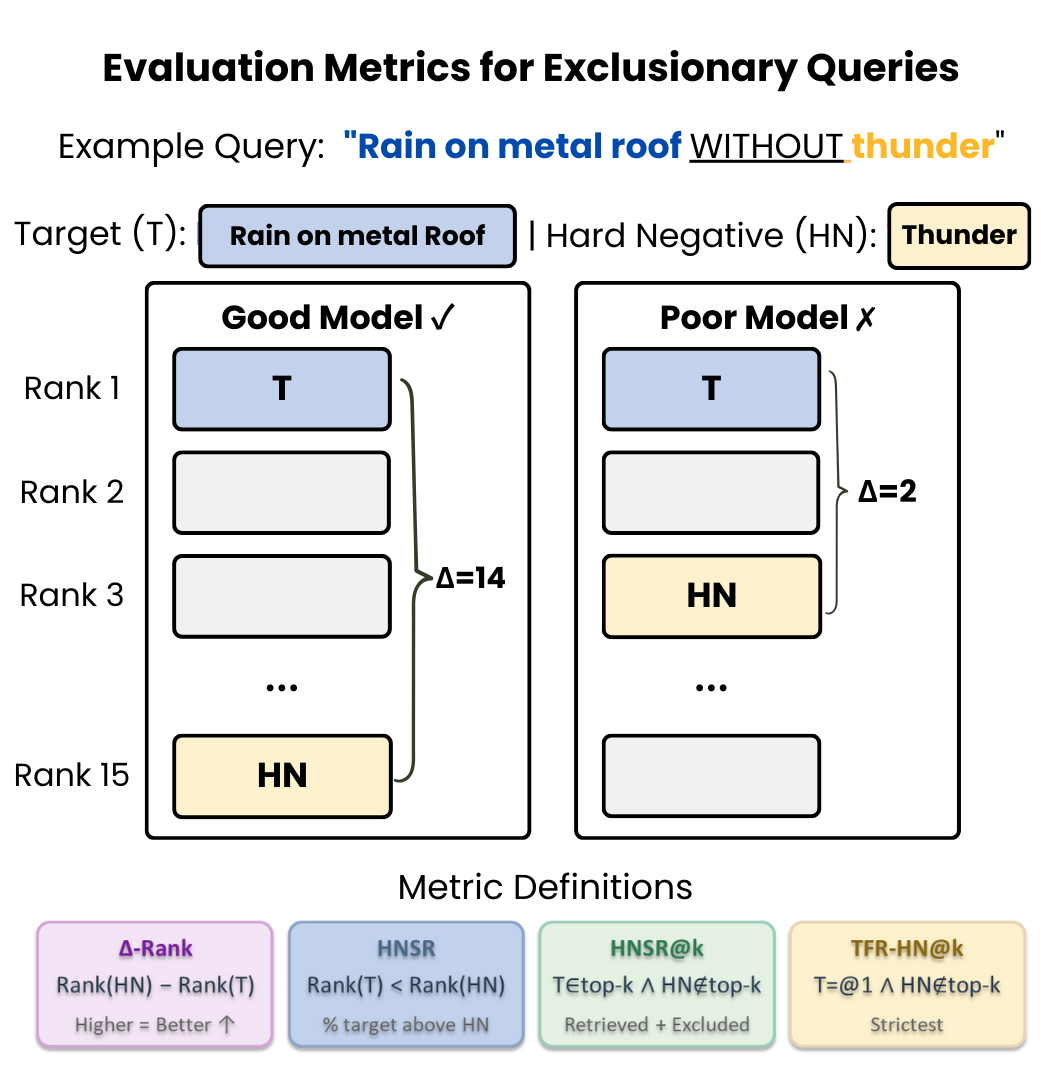}
\caption{Evaluation metrics for exclusionary queries. Standard R@k metrics only check if the Target audio is retrieved, but do not verify whether the model suppresses the Hard Negative (acoustically similar distractor). Our proposed HNSR@k explicitly measures the model's ability to rank the Target above the Hard Negative.}
\label{fig:exclusionary_metrics}
\end{figure}

Standard retrieval metrics (R@k) assess whether models retrieve relevant audio but do not measure whether models correctly suppress confusable alternatives. Figure~\ref{fig:exclusionary_metrics} illustrates this distinction. We introduce \textbf{HNSR@k} (Hard Negative Suppression Rate at $k$): the percentage of queries where the target is retrieved within top-$k$ AND the hard negative is ranked outside top-$k$. This metric captures both successful retrieval and distractor suppression---the core challenge of exclusionary queries. We also report \textbf{$\Delta$-Rank} = Rank(HN) $-$ Rank(T) as a continuous measure of target-distractor separation, where higher values indicate the model more effectively pushes hard negatives away from targets in the ranking. Additional metrics (TFR, TFR-HN@k) are detailed in Appendix~\ref{sec:additional_metrics}.

\begin{table*}[t]
\centering
\small
\begin{tabular}{l|ccc|ccc|ccc}
\toprule
\multirow{2}{*}{\textbf{Model}} & \multicolumn{3}{c|}{\textbf{AudioCaps}} & \multicolumn{3}{c|}{\textbf{Clotho}} & \multicolumn{3}{c}{\textbf{MECAT}} \\
& R@1 & R@5 & R@10 & R@1 & R@5 & R@10 & R@1 & R@5 & R@10 \\
\midrule
\multicolumn{10}{c}{\textit{CLAP Models}} \\
\midrule
LAION-CLAP & 22.15 & 54.63 & 69.93 & 13.34 & 33.49 & 46.22 & 6.80 & 20.93 & 30.66 \\
Robust-CLAP & 26.48 & 60.12 & 75.53 & 14.83 & 37.89 & 50.05 & 6.51 & 21.05 & 31.39 \\
MGA-CLAP & 36.39 & 70.11 & 82.46 & 18.78 & 43.79 & 56.63 & \textbf{8.02} & 23.51 & \uline{34.08} \\
M2D-CLAP & \textbf{41.39} & \textbf{77.13} & \textbf{88.27} & 17.55 & 42.91 & 55.54 & 7.37 & \uline{23.55} & 33.63 \\
\midrule
\multicolumn{10}{c}{\textit{Vanilla LALMs (No Retrieval Training)}} \\
\midrule
Nemotron-3B & 3.73 & 9.09 & 13.11 & 7.20 & 21.57 & 30.12 & 4.09 & 13.27 & 21.99 \\
Qwen2.5-Omni-3B & 0.37 & 1.19 & 1.87 & 0.17 & 0.63 & 1.13 & 0.12 & 0.63 & 1.57 \\
Qwen2.5-Omni-7B & 0.41 & 1.29 & 2.34 & 0.10 & 0.65 & 1.32 & 0.16 & 0.77 & 2.38 \\
\midrule
\multicolumn{10}{c}{\textit{OEA Models (Ours)}} \\
\midrule
OEA-Nemo3B & \uline{38.87} & \uline{72.64} & 83.88 & 19.04 & 40.57 & 54.24 & \uline{7.96} & \textbf{24.53} & \textbf{35.81} \\
OEA-Nemo3B (+Cl) & 35.75 & 68.64 & 81.35 & 21.57 & 47.16 & 60.36 & 6.98 & 23.00 & 33.41 \\
OEA-Qwen3B & 37.60 & 71.16 & 83.73 & 19.18 & 42.05 & 55.85 & 5.86 & 17.96 & 25.94 \\
OEA-Qwen3B (+Cl) & 35.96 & 69.35 & 81.99 & \textbf{22.87} & \textbf{49.78} & \textbf{63.25} & 5.78 & 17.16 & 24.76 \\
OEA-Qwen7B & 38.03 & 72.25 & \uline{84.53} & 19.77 & 44.78 & 57.65 & 7.74 & 23.29 & 33.49 \\
OEA-Qwen7B (+Cl) & 34.28 & 67.47 & 80.84 & \uline{22.53} & \uline{48.80} & \uline{62.70} & 7.02 & 22.94 & 33.12 \\
\bottomrule
\end{tabular}
\caption{Text-to-audio (T2A) baseline results (\%). M2D-CLAP achieves strong performance on AudioCaps, while OEA models excel on Clotho and MECAT. \textbf{Bold}: best per column, \uline{underline}: second best.}
\label{tab:baseline_t2a}
\end{table*}


\section{Experiments}

\subsection{Datasets}

\paragraph{Training Data}
OEA training follows a multi-stage curriculum: WavCaps (275,618 samples filtered to $\leq$31 seconds) for initial audio-text alignment, then AudioCaps v2 (91,256 training samples) for caption-based retrieval. Optionally, additional Clotho v2 training (3,839 clips) improves performance on natural audio descriptions. Models trained without Clotho are denoted as \textbf{OEA}; those with additional Clotho training are denoted \textbf{OEA (+Cl)}.

\paragraph{Evaluation Data}
We evaluate on three datasets: \textbf{AudioCaps} (975 clips, 5 captions each), \textbf{Clotho} (1,045 clips, 5 captions each), and \textbf{MECAT} (847 auto-captioned pairs).

A critical consideration for audio retrieval evaluation is \textbf{data leakage} between training and evaluation sets. WavCaps aggregates audio from multiple sources including AudioSet Strongly-Labeled, Freesound, BBC Sound Effects, and SoundBible. This creates potential overlap with evaluation benchmarks: AudioCaps derives from AudioSet, while Clotho sources from Freesound. Our analysis (Appendix~\ref{sec:leakage_analysis}) confirms significant overlap---17.7\% of AudioCaps test clips and 61.0\% of Clotho evaluation clips appear in WavCaps subsets.

To provide a \textbf{leakage-free evaluation}, we additionally evaluate on \textbf{MECAT} \cite{wu2025mecat}, which derives audio from ACAV100M \cite{lee2021acav100m}---a large-scale audio-visual dataset sourced from web videos that has no overlap with WavCaps training sources. While MECAT uses auto-generated captions (potentially noisier than human annotations in AudioCaps and Clotho), it provides an uncontaminated benchmark for assessing true generalization performance. Details of our leakage analysis are provided in Appendix~\ref{sec:leakage_analysis}.

\subsection{Models and Training}

\paragraph{Baselines}
We compare against four CLAP variants: LAION-CLAP \cite{wu2023large}, Robust-CLAP \cite{selvakumar2024robust}, MGA-CLAP \cite{li2024advancing}, and M2D-CLAP \cite{niizumi2025m2dclap}---the current state-of-the-art CLAP model that combines self-supervised masked modeling with contrastive language-audio pretraining. To isolate the contribution of our training approach, we also evaluate \textbf{vanilla LALMs}---Qwen2.5-Omni (3B and 7B) and Nemotron-3B without any retrieval-specific training. These baselines use mean-pooled last hidden states with L2 normalization, identical to OEA's embedding extraction.

\paragraph{OEA Variants}
We train six OEA variants across three backbones (Nemotron-3B, Qwen3B, Qwen7B): base OEA models and OEA (+Cl) variants with additional Clotho training. A discussion of scaling behavior and multi-backbone generality is provided in Appendix~\ref{sec:scaling_behavior}.

\paragraph{Training Configuration}
Training uses symmetric InfoNCE loss with $\tau=0.07$, AdamW optimizer (lr 3e-4 or 5e-4), and PyTorch DDP with BFloat16 precision. Training proceeds in stages with early stopping based on validation R@10.


\begin{table*}[t]
\centering
\small
\begin{tabular}{l|ccc|ccc|ccc}
\toprule
\multirow{2}{*}{\textbf{Model}} & \multicolumn{3}{c|}{\textbf{AudioCaps}} & \multicolumn{3}{c|}{\textbf{Clotho}} & \multicolumn{3}{c}{\textbf{MECAT}} \\
& R@1 & R@5 & R@10 & R@1 & R@5 & R@10 & R@1 & R@5 & R@10 \\
\midrule
\multicolumn{10}{c}{\textit{CLAP Models}} \\
\midrule
LAION-CLAP & 41.11 & 63.94 & 74.19 & 50.05 & 65.88 & 72.67 & 18.06 & 39.35 & 50.57 \\
Robust-CLAP & 41.09 & 63.96 & 74.22 & 50.05 & 65.88 & 72.67 & 18.06 & 39.35 & 50.57 \\
MGA-CLAP & 47.73 & 71.63 & 80.96 & 63.27 & 74.70 & 78.79 & 19.56 & 40.64 & 52.14 \\
M2D-CLAP & 47.51 & 70.03 & 79.96 & 55.85 & 69.05 & 74.76 & 17.92 & 38.74 & 49.51 \\
\midrule
\multicolumn{10}{c}{\textit{Vanilla LALMs (No Retrieval Training)}} \\
\midrule
Nemotron-3B & 33.17 & 52.27 & 62.13 & 57.84 & 69.36 & 74.26 & 6.86 & 20.77 & 29.56 \\
Qwen2.5-Omni-3B & 22.63 & 34.48 & 41.39 & 38.35 & 51.14 & 57.07 & 3.18 & 9.75 & 15.39 \\
Qwen2.5-Omni-7B & 24.06 & 37.19 & 44.62 & 40.23 & 54.26 & 60.27 & 3.16 & 10.46 & 15.92 \\
\midrule
\multicolumn{10}{c}{\textit{OEA Models (Ours)}} \\
\midrule
OEA-Nemo3B & 48.94 & \textbf{72.62} & \textbf{82.19} & 62.79 & 74.12 & 78.91 & 22.11 & 43.34 & 54.99 \\
OEA-Nemo3B (+Cl) & 48.94 & 72.21 & 81.64 & \uline{63.77} & \textbf{75.29} & \textbf{80.11} & 22.80 & 44.65 & 55.66 \\
OEA-Qwen3B & 49.42 & 72.27 & 81.76 & 62.81 & 73.76 & 78.22 & 22.60 & 45.38 & 56.60 \\
OEA-Qwen3B (+Cl) & 49.64 & 71.94 & 81.01 & \textbf{64.52} & \uline{75.25} & 79.71 & 24.65 & 46.23 & \uline{58.24} \\
OEA-Qwen7B & \textbf{50.30} & \uline{72.31} & \uline{82.05} & 63.66 & 74.32 & 79.37 & \uline{25.43} & \uline{47.19} & 58.18 \\
OEA-Qwen7B (+Cl) & \uline{49.93} & 71.79 & 81.05 & 63.58 & 75.04 & \uline{79.90} & \textbf{25.45} & \textbf{47.41} & \textbf{58.67} \\
\bottomrule
\end{tabular}
\caption{Text-to-text (T2T) baseline results (\%). OEA models substantially outperform all CLAP baselines including M2D-CLAP across all datasets. \textbf{Bold}: best per column, \uline{underline}: second best.}
\label{tab:baseline_t2t}
\end{table*}

\section{Results}

\subsection{Text-to-Audio Retrieval}

Table~\ref{tab:baseline_t2a} presents text-to-audio (T2A) retrieval results. Without retrieval-specific training, vanilla LALMs achieve only $\sim$1\% R@5, confirming that adaptation is essential for cross-modal retrieval.

On AudioCaps, M2D-CLAP achieves the strongest T2A performance (77.13\% R@5), while OEA models show competitive results (up to 72.64\% R@5). On the remaining benchmarks, OEA achieves comparable or superior performance: on Clotho, OEA-Qwen3B (+Cl) achieves best R@10 (63.25\%, +7.71 over M2D-CLAP), and on MECAT (leakage-free), OEA-Nemo3B achieves best R@5/R@10 (24.53\%/35.81\%). Overall, \textbf{T2A performance is similar across methods}, with M2D-CLAP excelling on in-domain AudioCaps and OEA demonstrating stronger cross-domain generalization.

\subsection{Caption-Based Text-to-Text Retrieval}

We evaluate \textbf{text-to-text (T2T)} retrieval (Table~\ref{tab:baseline_t2t}), simulating caption-based audio retrieval where audio is first converted to text via automatic captioning \cite{chu2024qwenaudio, xu2025qwen25omni}. T2T achieves substantially higher performance than T2A (e.g., 47.41\% vs. 22.94\% R@5 on MECAT), suggesting caption-based pipelines offer a practical alternative.

\textbf{OEA Dominates T2T.} OEA models substantially outperform all CLAP baselines: +8.67 R@1 over M2D-CLAP on Clotho (64.52\% vs. 55.85\%) and +22\% relative improvement on MECAT R@5 (47.41\% vs. 38.74\%). This dominance stems from OEA's unified LLM backbone processing both query and caption text, enabling richer semantic matching compared to CLAP's separate and smaller text encoders. While CLAP models use lightweight text encoders optimized for contrastive alignment, OEA leverages the full representational capacity of billion-parameter LLMs for text understanding. This suggests that for caption-based retrieval pipelines---increasingly practical with modern audio captioning systems---LLM-based encoders offer substantial advantages over traditional CLAP architectures.

\subsection{User-Intent Query to Audio Retrieval}

\subsubsection{Conversational and Reformulation Queries}

Table~\ref{tab:uiq_mean} presents UIQ performance across Question, Imperative, Keyphrase, and Paraphrase query types.

\begin{table*}[t]
\centering
\small
\begin{tabular}{l|cc|cc|c|c}
\toprule
\multirow{2}{*}{\textbf{Model}} & \multicolumn{2}{c|}{\textbf{Conversational}} & \multicolumn{2}{c|}{\textbf{Reformulation}} & \textbf{Exclusionary} & \multirow{2}{*}{\textbf{Avg UIQ}} \\
& Question & Imperative & Keyphrase & Paraphrase & Hard Neg. & \\
\midrule
LAION-CLAP & 38.65 & 37.06 & 42.71 & 40.92 & 24.7 & 38.01 \\
Robust-CLAP & 43.24 & 41.07 & 47.19 & 43.98 & 30.3 & 42.07 \\
MGA-CLAP & 44.41 & 44.94 & \uline{50.97} & 48.74 & 21.2 & 45.16 \\
M2D-CLAP & \textbf{48.76} & 44.74 & \textbf{53.16} & \textbf{50.58} & 30.3 & \textbf{47.76} \\
\midrule
OEA-Nemo3B & 46.65 & 47.58 & 48.87 & 49.19 & \uline{32.9} & 46.39 \\
OEA-Nemo3B (+Cl) & 46.10 & \uline{49.41} & 49.02 & 48.82 & 32.5 & 46.97 \\
OEA-Qwen3B & 44.17 & 46.06 & 48.20 & 46.94 & 31.6 & 44.57 \\
OEA-Qwen3B (+Cl) & 46.71 & 48.55 & 50.07 & 49.05 & 32.8 & \uline{47.18} \\
OEA-Qwen7B & \uline{47.47} & 49.30 & 50.31 & 50.12 & 31.8 & 47.08 \\
OEA-Qwen7B (+Cl) & 47.32 & \textbf{49.87} & 50.21 & \uline{50.25} & \textbf{34.6} & 47.16 \\
\bottomrule
\end{tabular}
\caption{Mean UIQ results across AudioCaps, Clotho, and MECAT (\%). Queries are organized by category: Conversational (Question, Imperative), Reformulation (Keyphrase, Paraphrase), and Exclusionary (Hard Neg.). Conversational, Reformulation, and Avg UIQ use R@5; Exclusionary uses HNSR@10 to measure hard negative discrimination ability. M2D-CLAP achieves best Avg UIQ, while OEA models achieve best Imperative and Hard Negative discrimination. \textbf{Bold}: best, \uline{underline}: second best.}
\label{tab:uiq_mean}
\end{table*}

\paragraph{UIQ Performance: Comparable with Complementary Strengths}
For standard UIQ retrieval, M2D-CLAP and OEA show \textbf{comparable overall performance} with complementary strengths. M2D-CLAP achieves the highest mean scores on Question (48.76\%), Keyphrase (53.16\%), and Paraphrase (50.58\%) queries, while OEA models achieve the best performance on \textbf{Imperative} queries---the most command-like formulation. OEA-Qwen7B (+Cl) achieves 49.87\% compared to M2D-CLAP's 44.74\%---a +5.13 point gain (+11.5\% relative), confirming that LLM backbones excel at parsing command-style queries. M2D-CLAP achieves the best overall \textbf{Avg UIQ} (47.76\%) with OEA-Qwen3B (+Cl) achieving competitive second place (47.18\%, -0.58\%p). Notably, OEA's key advantages emerge in \textbf{T2T retrieval} (+22\% improvement) and \textbf{hard negative discrimination} (detailed below), rather than standard UIQ performance.

\subsubsection{Exclusionary Query Evaluation}

Negative queries represent the most challenging UIQ type, requiring models to understand both desired content and explicit exclusions. As shown in Table~\ref{tab:uiq_mean}, the Exclusionary category reports HNSR@10 to measure hard negative discrimination ability.

\paragraph{Standard Retrieval vs. Discrimination}
M2D-CLAP achieves the best standard retrieval on exclusionary queries (41.56\% R@5), contributing to its highest overall Avg UIQ. However, \textbf{standard metrics do not capture discrimination ability}---the core challenge where models must distinguish between acoustically similar but semantically distinct audio.

\paragraph{OEA Dominates Hard Negative Discrimination}
OEA-Qwen7B (+Cl) achieves the best HNSR@10 (34.6\% vs. M2D-CLAP's 30.3\%, +4.3\%p), demonstrating superior hard negative discrimination. Additional metrics in Appendix~\ref{sec:negative_appendix} confirm this pattern: OEA achieves HNSR 74.4\% vs. M2D-CLAP's 68.0\% (+6.4\%p), and TFR@10 10.1\% vs. 7.5\% (+34.7\% relative improvement).

\paragraph{Implications}
These results reveal a critical distinction: while T2A retrieval and standard UIQ performance show \textbf{comparable results between M2D-CLAP and OEA}, the key differentiators emerge in two areas where OEA demonstrates clear superiority: (1) \textbf{T2T retrieval}, where OEA achieves +22\% relative improvement leveraging its LLM backbone for richer text-to-text semantic matching, and (2) \textbf{hard negative discrimination}, where OEA substantially outperforms (+4.3\%p HNSR@10, +34.7\% relative TFR@10). For exclusionary queries where users explicitly specify what they \textit{don't} want, discrimination ability is arguably more important than raw retrieval performance. To move beyond a monolithic ``LLM understanding'' explanation, we conduct mechanism ablations on the released OEA-Qwen7B (+Cl) checkpoint (Appendix~\ref{sec:mechanism_ablations}). They reveal a \textit{joint} mechanism: (1) OEA's \textbf{audio embeddings} place targets and hard negatives significantly farther apart than M2D-CLAP's on the same pairs (all $p \ll 0.001$), so confusable clips are separable before any query is applied; and (2) the exclusion signal is carried by \textbf{attention over query structure} rather than the negation word itself---shuffling query word order (identical tokens, ``without'' preserved) collapses exclusionary separation ($\Delta$-Rank $22.5 \to -7.2$) while leaving positive queries essentially unaffected. A cross-model control applying the identical shuffle to the CLAP baselines shows, however, that this order sensitivity is \textit{general}: every text encoder collapses on exclusionary queries (all $p \ll 10^{-3}$), and OEA's exclusion-specific structure contribution sits between the baselines' (above M2D and MGA, below LAION). The decisive, OEA-specific factor is therefore the audio-side separation in (1)---the magnitude of separation OEA achieves---rather than a text-side mechanism CLAP lacks. Notably, the negation clause is not converted into a clean downward push on the excluded clip: stripping the exclusion clause improves suppression for \textit{every} model, because the clause lexically mentions the excluded content; OEA is simply the most robust to this lexical pull (smallest HNSR@10 degradation).


\section{Conclusion}

We presented \textbf{Omni-Embed-Audio (OEA)}, a unified encoder leveraging multimodal LLMs for audio-text retrieval. For text-to-audio (T2A) retrieval, OEA achieves \textbf{comparable performance} to M2D-CLAP overall, with stronger cross-domain generalization on Clotho and MECAT. OEA's key advantages emerge in two critical areas: (1) \textbf{Text-to-text (T2T) retrieval}, where OEA \textit{substantially outperforms} all baselines (+8.67\%p R@1 on Clotho, +22\% relative improvement on MECAT), demonstrating that LLM backbones provide superior text-to-text semantic matching for caption-based retrieval pipelines; and (2) \textbf{Hard negative discrimination}, where OEA achieves +4.3\%p HNSR@10 and +34.7\% relative TFR@10 improvement; mechanism ablations (Appendix~\ref{sec:mechanism_ablations}) attribute this chiefly to more separable audio embeddings, with order-sensitive processing of exclusion structure that cross-model controls show is shared with CLAP encoders. We introduced the \textbf{UIQ benchmark} with five query types and novel \textbf{discrimination metrics} (HNSR, TFR), exposing that standard retrieval metrics fail to capture models' ability to distinguish acoustically similar but semantically distinct audio---a critical capability for real-world search applications.


\section*{Limitations}

Our work has several limitations. First, OEA requires multimodal LLM backbones with native audio understanding, limiting base model choices; adopting text-only LLMs would require a separate audio encoder, which alters the unified encoder design. While we validate across three backbones from two organizations (Appendix~\ref{sec:scaling_behavior}), extending to additional architectures remains an important direction. Second, the OEA architecture incurs higher computational cost than lightweight CLAP models, particularly for audio encoding (Appendix~\ref{sec:efficiency_analysis}); however, per-query text latency remains practical for real-time serving, and quantization can substantially reduce memory requirements. Third, hard negative mining relies on MGA-CLAP for acoustic candidate retrieval, potentially omitting confusion types that MGA-CLAP fails to identify; we mitigate this through human-guided calibration of the mining thresholds on sampled pairs (Appendix~\ref{sec:hn_human_validation}) and dual-model filtering (acoustic + semantic). Fourth, UIQ generation uses a single LLM (GPT-5.1), potentially introducing systematic biases; our analysis (Appendix~\ref{sec:uiq_bias_analysis}) suggests that OEA does not uniformly benefit from shared linguistic patterns, but future work should validate against real-world user query distributions. Finally, retrieval performance saturates at the 3B scale, with the 7B model offering marginal gains on some metrics (Appendix~\ref{sec:scaling_behavior}), suggesting that the bottleneck shifts from backbone capacity to contrastive alignment quality at larger scales.


\section*{Acknowledgments}

This work was partly supported by the Institute of Information \& Communications Technology Planning \& Evaluation (IITP) grant funded by the Korea government (MSIT) (No.~RS-2025-25441313, Professional AI Talent Development Program for Multimodal AI Agents, Contribution: 50\%). This research was also supported by the MSIT, Korea, under the Top-Tier AI Global HRD invitation program (RS-2025-25461932) supervised by the IITP.

\bibliography{custom}


\appendix

\section{Implementation Details}
\label{sec:implementation_details}

\paragraph{Input Processing}
For text encoding, queries are wrapped in a chat template with a \texttt{query:} prefix, tokenized, passed through the transformer, and mean-pooled over the last hidden layer. For audio encoding, raw waveforms are loaded at 16kHz mono, converted to audio features via the model's native audio processor, wrapped with a \texttt{passage:} prefix, and processed through the same transformer with identical mean pooling.

\paragraph{Adaptation Layers}
LoRA is configured with rank $r=16$, $\alpha=32$, and dropout 0.05, targeting query, key, value, and output projections of all attention layers. Each modality has a dedicated projection head consisting of a bias-free linear layer (hidden dim $\rightarrow$ 512), dropout (0.1), layer normalization, and L2 normalization to produce unit-norm embeddings.

\paragraph{Training Objective}
We use symmetric contrastive learning with InfoNCE loss:
\begin{equation}
\mathcal{L} = \frac{1}{2}\left(\mathcal{L}_{t \rightarrow a} + \mathcal{L}_{a \rightarrow t}\right)
\end{equation}
where $\mathcal{L}_{t \rightarrow a} = -\log \frac{\exp(\text{sim}(t_i, a_i)/\tau)}{\sum_j \exp(\text{sim}(t_i, a_j)/\tau)}$ and $\tau = 0.07$ is the temperature parameter.

\section{Data Leakage Analysis}
\label{sec:leakage_analysis}

A critical but often overlooked issue in audio retrieval evaluation is \textbf{data leakage} between training corpora and evaluation benchmarks. WavCaps \cite{mei2024wavcaps}, a widely-used pretraining dataset, aggregates audio from multiple sources that overlap with standard evaluation benchmarks. We conduct a systematic analysis of potential contamination.

\subsection{Dataset Provenance}

Table~\ref{tab:dataset_sources} summarizes the source relationships between training and evaluation datasets.

\begin{table}[h]
\centering
\footnotesize
\begin{tabular}{ll}
\toprule
\textbf{Evaluation Set} & \textbf{Original Source} \\
\midrule
AudioCaps & AudioSet \\
Clotho & Freesound \\
SoundDescs & BBC Sound Effects \\
FSD50K & Freesound \\
ESC-50 & Freesound \\
DCASE2022 & Freesound \\
\midrule
MECAT & ACAV100M (web videos) \\
\bottomrule
\end{tabular}
\caption{Source provenance of evaluation datasets. WavCaps includes AudioSet Strongly-Labeled, Freesound, BBC Sound Effects, and SoundBible subsets, creating potential overlap with most standard benchmarks except MECAT.}
\label{tab:dataset_sources}
\end{table}

\subsection{AudioCaps--WavCaps Overlap}

We analyzed overlap between the AudioCaps test set (4,875 caption rows covering 975 unique YouTube IDs) and the WavCaps AudioSet\_SL subset (108,317 captioned clips stored as \texttt{Y<YouTubeID>.wav}). We normalized every WavCaps ID by stripping the leading `Y' and trailing `.wav' to restore the raw 11-character YouTube ID, then matched against AudioCaps test metadata.

\paragraph{Findings}
\begin{itemize}
    \item \textbf{173 of 975} unique AudioCaps test clips (17.7\%) are present in WavCaps AudioSet\_SL (representing 0.16\% of WavCaps entries)
    \item Because AudioCaps supplies five captions per clip, this manifests as \textbf{865} duplicated caption rows
    \item Example overlaps include: \texttt{6BJ455B1aAs} (rocket/missile launch with explosion), \texttt{VjSEIRnLAh8} (frying food with female speech), \texttt{ztSjcZNUY7A} (crying baby with woman speaking)
    \item Captions on both sides describe identical acoustic content, confirming shared underlying audio rather than accidental ID collisions
\end{itemize}

\subsection{Clotho--WavCaps Overlap}

We analyzed overlap between the Clotho v2 evaluation split (1,045 audio files with five captions each) and the WavCaps Freesound subset. We normalized Clotho's \texttt{file\_name} field and matched it case-insensitively against filenames in the Freesound metadata JSON.

\paragraph{Findings}
\begin{itemize}
    \item \textbf{638 of 1,045} Clotho evaluation clips (61.0\%) have identical filenames in WavCaps Freesound metadata, proving they are the exact same recordings (e.g., \texttt{Radio Garble.wav} $\leftrightarrow$ Freesound ID 80399)
    \item Captions in WavCaps often mention the same acoustic events (e.g., ``Simulated garbled radio traffic with crosstalk''), confirming this is not merely a naming collision
    \item The remaining 39\% likely still originate from Freesound but were renamed during Clotho curation; additional fuzzy matching (author + duration or audio fingerprinting) would surface more overlaps
\end{itemize}

\subsection{Implications and Mitigation}

These findings have significant implications for audio retrieval evaluation:

\begin{itemize}
    \item Any model trained on WavCaps AudioSet\_SL without filtering overlapping IDs will have memorized nearly one-fifth of the AudioCaps test set, invalidating downstream evaluations
    \item Similarly, models trained on unfiltered WavCaps Freesound have already ``seen'' the majority of Clotho evaluation audio; reported Clotho performance is therefore inflated
    \item Prior experiments that mixed these datasets should be considered contaminated unless they explicitly removed overlapping samples
\end{itemize}

We apply blocklists based on our overlap analysis to exclude contaminated samples from training. To provide an uncontaminated evaluation baseline, we include \textbf{MECAT} \cite{wu2025mecat}, which derives from ACAV100M \cite{lee2021acav100m}---a dataset of web videos with no overlap with WavCaps sources. Our embedding-based and filename-based leakage checks found no significant overlap between MECAT and WavCaps. While MECAT uses auto-generated captions (potentially noisier than human annotations), it provides the only truly leakage-free benchmark among our evaluation sets.

\section{UIQ Benchmark Validation}
\label{sec:uiq_validation_appendix}

To ensure UIQ quality, we validate through LLM-based evaluation (Claude Opus 4.5) and human annotation.

\paragraph{5-Point Likert Scale Definitions}
\begin{itemize}
    \item \textbf{1 - Incompatible}: Query meaning completely diverges from original caption or target intent
    \item \textbf{2 - Poor}: Query captures some elements but introduces significant semantic drift
    \item \textbf{3 - Acceptable}: Query roughly conveys the intended meaning with minor issues
    \item \textbf{4 - Good}: Query accurately reflects original meaning/intent in target format
    \item \textbf{5 - Excellent}: Query perfectly preserves semantics with natural formulation
\end{itemize}

\begin{table}[h]
\centering
\footnotesize
\setlength{\tabcolsep}{3.5pt}
\begin{tabular}{l|ccc|ccc}
\toprule
\multirow{2}{*}{\textbf{Query Type}} & \multicolumn{3}{c|}{\textbf{Per Dataset Mean}} & \multicolumn{3}{c}{\textbf{Overall}} \\
& AC & Cl & ME & Mean & Std & N \\
\midrule
Question & 4.11 & 4.13 & 4.53 & 4.26 & 0.91 & 135 \\
Imper. & 4.31 & 3.64 & 4.53 & 4.16 & 1.06 & 135 \\
Keyph. & 4.40 & 4.02 & 4.62 & 4.35 & 1.05 & 135 \\
Paraph. & 4.42 & 3.69 & 4.31 & 4.14 & 1.21 & 135 \\
Negative & 3.73 & 3.87 & 3.87 & 3.82 & 1.25 & 135 \\
\midrule
\textbf{Mean} & 4.20 & 3.87 & 4.37 & \textbf{4.15} & 1.12 & 675 \\
\bottomrule
\end{tabular}
\caption{Detailed UIQ human evaluation results by dataset. AC: AudioCaps, Cl: Clotho, ME: MECAT. 9 annotators rated 15 samples (5 per dataset) across 5 UIQ types. MECAT achieves highest validity (4.37), likely due to cleaner auto-generated captions. Clotho shows lower scores (3.87), reflecting ambiguity in some natural audio descriptions.}
\label{tab:uiq_validation_appendix}
\end{table}


\section{UIQ Generation Prompts}
\label{sec:uiq_prompts}

This section documents the prompts used to generate User-Intent Queries (UIQs) using GPT-5.1.

\subsection{System Prompt}

The following system prompt instructs the model to generate all five UIQ types in a single API call:

\begin{lstlisting}
You generate five retrieval queries for one audio example 
(with a target and a negative).
Return exactly FIVE lines, one per type, in the form:
question: <text>
imperative: <text>
paraphrase: <text>
tagging: <text>
negative: <text>

Type-specific rules:
- question: Natural question form, 8-18 words; 
  starts with Can you/Do you/Are there/Is there; ends with ?.
- imperative: Direct command, 8-15 words; 
  starts with Find/Search for/Locate/Retrieve; no ?.
- paraphrase: 12-25 word declarative sentence; 
  no command/question tone.
- tagging: 3-6 comma-separated tags; each 1-4 words; lowercase.
- negative: 15-35 words; clearly state desired sounds AND 
  exclusions using without/not/excluding.

Global rules:
- One line per type, in the exact order above.
- No numbering, bullets, or extra commentary.
\end{lstlisting}

\subsection{User Prompt Template}

For each audio sample, the following template is populated:

\begin{lstlisting}
TARGET audio id: {target_audio_id}
NEGATIVE audio id: {negative_audio_id}

TARGET descriptions:
- {target_caption_1}
- {target_caption_2}

NEGATIVE descriptions:
- {negative_caption_1}
- {negative_caption_2}

Generate all five query types as described.
\end{lstlisting}

\subsection{Generation Parameters}

\begin{table}[h]
\centering
\small
\begin{tabular}{l|l}
\toprule
\textbf{Parameter} & \textbf{Value} \\
\midrule
Model & GPT-5.1 \\
Temperature & 0.35 \\
Top-p & 0.9 \\
Max tokens & 256 \\
\bottomrule
\end{tabular}
\caption{UIQ generation parameters.}
\label{tab:uiq_gen_params}
\end{table}

\subsection{Generated Query Examples}

Table~\ref{tab:uiq_examples} shows example UIQs generated for a sample audio clip.

\begin{table}[h]
\centering
\footnotesize
\begin{tabular}{l|p{5.5cm}}
\toprule
\textbf{Type} & \textbf{Generated Query} \\
\midrule
Question & Can you find clear dog barks echoing in a large hall? \\
\midrule
Imperative & Find crisp footsteps on gravel with light echo \\
\midrule
Paraphrase & Echoing dog barks resonate through a large empty hall \\
\midrule
Keyphrase & dog barks, echoing hall, reverberant \\
\midrule
Negative & Heavy rain and wind on metal surfaces without thunder or engine noise \\
\bottomrule
\end{tabular}
\caption{Example UIQs for different query types.}
\label{tab:uiq_examples}
\end{table}


\section{Human and LLM Evaluation Prompts}
\label{sec:eval_prompts}

This section documents the evaluation methodology for validating UIQ quality.

\subsection{Human Evaluation}

\paragraph{Evaluation Question}
The web-based listening test presents annotators with the following question:

\begin{quote}
\textit{``Validity -- Does this make sense as a search query regarding the original captions? If needed, please check the audio too.''}
\end{quote}

\paragraph{Rating Scale}
Annotators rate queries on a 5-point Likert scale:
\begin{itemize}
    \item \textbf{1 (Poor)}: Completely invalid or nonsensical
    \item \textbf{2}: Major issues, barely related
    \item \textbf{3 (Average)}: Acceptable with some issues
    \item \textbf{4}: Good, appropriately reflects the caption
    \item \textbf{5 (Excellent)}: Perfect validity and natural formulation
\end{itemize}

\paragraph{Interface}
Each sample presents: (1) audio player for target audio, (2) original human-written captions, (3) five generated UIQs (one per type), (4) rating buttons (1--5) for each UIQ, and (5) optional comment field.

\subsection{LLM Evaluation}

We employ Claude Opus 4.5 for automated evaluation using the same validity criterion as human evaluation, enabling direct comparison via KL-divergence.

\subsubsection{Evaluation Prompt}

\begin{lstlisting}
You are evaluating a generated search query for an audio 
retrieval system.

**Original Audio Caption:** {caption}
**Generated Query:** {query}
**Query Type:** {query_type}

**Task:** Rate the validity on a scale of 1-5.

**Question:** Does this make sense as a search query 
regarding the original caption?

**Rating Scale:**
- 1: Completely invalid or nonsensical
- 2: Poor - major issues, barely related
- 3: Acceptable - roughly conveys intent with issues
- 4: Good - valid, appropriately reflects caption
- 5: Excellent - perfectly captures intent naturally

Respond with ONLY: {"score": <1-5>, "reasoning": "<brief>"}
\end{lstlisting}

\subsubsection{LLM Evaluation Parameters}

\begin{table}[h]
\centering
\small
\begin{tabular}{l|l}
\toprule
\textbf{Parameter} & \textbf{Value} \\
\midrule
Model & Claude Opus 4.5 \\
Max tokens & 256 \\
Rate limiting & 0.5s between requests \\
\bottomrule
\end{tabular}
\caption{LLM evaluation parameters.}
\label{tab:llm_eval_params}
\end{table}


\section{Detailed Baseline Results}
\label{sec:detailed_baseline}

This section provides mean results across datasets for baseline caption query evaluation.

\begin{table*}[h]
\centering
\small
\begin{tabular}{l|ccc|ccc}
\toprule
\multirow{2}{*}{\textbf{Model}} & \multicolumn{3}{c|}{\textbf{T2A}} & \multicolumn{3}{c}{\textbf{T2T}} \\
& R@1 & R@5 & R@10 & R@1 & R@5 & R@10 \\
\midrule
\multicolumn{7}{c}{\textit{CLAP Models}} \\
\midrule
LAION-CLAP & 14.10 & 36.35 & 48.94 & 36.41 & 56.39 & 65.81 \\
Robust-CLAP & 15.94 & 39.69 & 52.32 & 36.40 & 56.39 & 65.82 \\
MGA-CLAP & 21.06 & 45.80 & 57.72 & 43.52 & 62.32 & 70.63 \\
M2D-CLAP & \textbf{22.10} & \textbf{47.86} & \textbf{59.15} & 40.43 & 59.27 & 68.08 \\
\midrule
\multicolumn{7}{c}{\textit{Vanilla LALMs}} \\
\midrule
Nemotron-3B & 5.00 & 14.64 & 21.74 & 32.62 & 47.47 & 55.32 \\
Qwen2.5-Omni-3B & 0.22 & 0.82 & 1.52 & 21.39 & 31.79 & 37.95 \\
Qwen2.5-Omni-7B & 0.22 & 0.90 & 2.01 & 22.49 & 33.97 & 40.27 \\
\midrule
\multicolumn{7}{c}{\textit{OEA Models}} \\
\midrule
OEA-Nemo3B & 21.96 & 45.91 & 57.98 & 44.62 & 63.36 & 72.03 \\
OEA-Nemo3B (+Cl) & 21.43 & 46.26 & 58.38 & 45.17 & 64.05 & 72.47 \\
OEA-Qwen3B & 20.88 & 43.72 & 55.17 & 44.94 & 63.80 & 72.20 \\
OEA-Qwen3B (+Cl) & 21.54 & 45.43 & 56.67 & 46.27 & 64.47 & 72.98 \\
OEA-Qwen7B & \uline{21.85} & \uline{46.77} & 58.56 & \textbf{46.46} & \uline{64.60} & \uline{73.20} \\
OEA-Qwen7B (+Cl) & 21.27 & 46.40 & \uline{58.89} & \uline{46.32} & \textbf{64.75} & \textbf{73.21} \\
\bottomrule
\end{tabular}
\caption{Mean baseline results across AudioCaps, Clotho, and MECAT (\%). M2D-CLAP achieves best T2A performance, while OEA models substantially outperform all baselines on T2T. \textbf{Bold}: best per column, \uline{underline}: second best.}
\label{tab:baseline_mean}
\end{table*}


\section{Detailed UIQ Results}
\label{sec:detailed_uiq}

This section provides complete per-dataset results for each UIQ type.

\subsection{Question Query Results}

\begin{table*}[h]
\centering
\small
\begin{tabular}{l|ccc|ccc|ccc}
\toprule
\multirow{2}{*}{\textbf{Model}} & \multicolumn{3}{c|}{\textbf{AudioCaps}} & \multicolumn{3}{c|}{\textbf{Clotho}} & \multicolumn{3}{c}{\textbf{MECAT}} \\
& R@1 & R@5 & R@10 & R@1 & R@5 & R@10 & R@1 & R@5 & R@10 \\
\midrule
LAION-CLAP & 25.13 & 56.10 & 71.18 & 16.17 & 39.33 & 53.30 & 6.84 & 20.52 & 31.25 \\
Robust-CLAP & 26.15 & 63.28 & 77.13 & 18.09 & 44.02 & 57.03 & 7.43 & 22.41 & 31.96 \\
MGA-CLAP & 31.90 & 65.03 & 78.87 & 18.18 & 44.98 & 57.70 & 8.25 & 23.23 & \uline{36.20} \\
M2D-CLAP & \textbf{37.13} & \textbf{73.74} & \textbf{86.56} & 20.00 & 46.60 & 58.66 & \textbf{9.43} & \textbf{25.94} & \textbf{37.62} \\
\midrule
OEA-Nemo3B & 36.72 & \uline{72.00} & 83.69 & 19.81 & 43.54 & 57.61 & 7.31 & \uline{24.41} & 35.14 \\
OEA-Nemo3B (+Cl) & 34.56 & 65.64 & 80.31 & 23.44 & 50.62 & 63.73 & 6.96 & 22.05 & 32.43 \\
OEA-Qwen3B & 36.72 & 70.36 & 82.56 & 21.34 & 44.11 & 59.71 & 7.43 & 18.04 & 27.12 \\
OEA-Qwen3B (+Cl) & 35.18 & 68.31 & 81.74 & \textbf{25.74} & \textbf{54.26} & \textbf{66.79} & 6.60 & 17.57 & 24.88 \\
OEA-Qwen7B & \uline{36.92} & 70.26 & \uline{84.72} & 22.30 & 48.33 & 63.06 & \uline{8.37} & 23.82 & 34.79 \\
OEA-Qwen7B (+Cl) & 33.23 & 66.56 & 79.08 & \uline{24.11} & \uline{52.15} & \uline{66.32} & 8.02 & 23.23 & 34.79 \\
\bottomrule
\end{tabular}
\caption{Question query T2A results (\%). \textbf{Bold}: best per column, \uline{underline}: second best.}
\label{tab:question_detailed}
\end{table*}

\subsection{Imperative Query Results}

\begin{table*}[h]
\centering
\small
\begin{tabular}{l|ccc|ccc|ccc}
\toprule
\multirow{2}{*}{\textbf{Model}} & \multicolumn{3}{c|}{\textbf{AudioCaps}} & \multicolumn{3}{c|}{\textbf{Clotho}} & \multicolumn{3}{c}{\textbf{MECAT}} \\
& R@1 & R@5 & R@10 & R@1 & R@5 & R@10 & R@1 & R@5 & R@10 \\
\midrule
LAION-CLAP & 22.36 & 55.38 & 70.05 & 14.45 & 37.99 & 51.29 & 5.66 & 17.81 & 27.24 \\
Robust-CLAP & 26.77 & 62.36 & 77.54 & 16.08 & 42.58 & 56.94 & 5.66 & 18.28 & 28.77 \\
MGA-CLAP & 34.67 & 67.59 & 79.90 & 17.99 & 44.59 & 55.79 & 8.25 & 22.64 & 33.14 \\
M2D-CLAP & 33.85 & 67.49 & 80.10 & 20.29 & 46.22 & 60.00 & 7.08 & 20.52 & 31.25 \\
\midrule
OEA-Nemo3B & \uline{41.33} & \textbf{75.08} & \textbf{86.36} & 18.85 & 43.25 & 57.03 & \uline{8.37} & \textbf{24.41} & \textbf{36.44} \\
OEA-Nemo3B (+Cl) & 37.64 & 72.51 & 83.59 & 24.02 & 51.29 & 64.69 & 7.67 & \uline{24.41} & 34.55 \\
OEA-Qwen3B & 40.00 & 72.72 & \uline{84.92} & 22.30 & 46.12 & 59.33 & 7.43 & 19.34 & 27.00 \\
OEA-Qwen3B (+Cl) & 39.49 & 72.62 & 84.31 & \uline{25.45} & \textbf{55.69} & \textbf{67.56} & 7.19 & 17.33 & 25.12 \\
OEA-Qwen7B & \textbf{42.15} & \uline{74.77} & 84.82 & 24.02 & 48.71 & 61.82 & \textbf{8.49} & 24.41 & 34.55 \\
OEA-Qwen7B (+Cl) & 36.72 & 70.56 & 83.18 & \textbf{26.22} & \uline{54.64} & \uline{67.27} & 8.02 & 24.41 & \uline{35.61} \\
\bottomrule
\end{tabular}
\caption{Imperative query T2A results (\%). \textbf{Bold}: best per column, \uline{underline}: second best.}
\label{tab:imperative_detailed}
\end{table*}

\subsection{Paraphrase Query Results}

\begin{table*}[h]
\centering
\small
\begin{tabular}{l|ccc|ccc|ccc}
\toprule
\multirow{2}{*}{\textbf{Model}} & \multicolumn{3}{c|}{\textbf{AudioCaps}} & \multicolumn{3}{c|}{\textbf{Clotho}} & \multicolumn{3}{c}{\textbf{MECAT}} \\
& R@1 & R@5 & R@10 & R@1 & R@5 & R@10 & R@1 & R@5 & R@10 \\
\midrule
LAION-CLAP & 28.21 & 62.97 & 77.95 & 15.98 & 39.14 & 52.73 & 6.49 & 20.64 & 30.78 \\
Robust-CLAP & 30.97 & 67.90 & 83.28 & 18.28 & 42.58 & 56.94 & 6.25 & 21.46 & 31.37 \\
MGA-CLAP & 38.87 & 73.54 & 87.28 & 19.62 & 47.56 & 59.71 & \textbf{9.43} & \uline{25.12} & \textbf{35.97} \\
M2D-CLAP & \uline{42.36} & \uline{76.62} & \textbf{88.92} & 21.63 & 50.72 & 60.67 & 7.43 & 24.41 & 34.55 \\
\midrule
OEA-Nemo3B & 41.85 & \textbf{77.85} & \uline{88.00} & 20.96 & 44.50 & 57.22 & 7.43 & \textbf{25.24} & 35.85 \\
OEA-Nemo3B (+Cl) & 38.05 & 73.33 & 85.13 & 23.83 & 50.14 & 64.50 & 7.67 & 23.00 & 33.96 \\
OEA-Qwen3B & 40.10 & 74.56 & 86.05 & 21.24 & 46.79 & 60.19 & 6.49 & 19.46 & 26.89 \\
OEA-Qwen3B (+Cl) & 39.18 & 73.03 & 84.92 & \textbf{27.56} & \textbf{55.50} & \textbf{70.81} & 5.90 & 18.63 & 25.59 \\
OEA-Qwen7B & \textbf{42.67} & 75.90 & 87.69 & 22.78 & 50.53 & 61.82 & \uline{9.08} & 23.94 & 35.02 \\
OEA-Qwen7B (+Cl) & 38.56 & 72.10 & 84.10 & \uline{26.41} & \uline{54.83} & \uline{67.37} & 7.31 & 23.82 & \uline{35.97} \\
\bottomrule
\end{tabular}
\caption{Paraphrase query T2A results (\%). \textbf{Bold}: best per column, \uline{underline}: second best.}
\label{tab:paraphrase_detailed}
\end{table*}

\subsection{Keyphrase Query Results}

\begin{table*}[h]
\centering
\small
\begin{tabular}{l|ccc|ccc|ccc}
\toprule
\multirow{2}{*}{\textbf{Model}} & \multicolumn{3}{c|}{\textbf{AudioCaps}} & \multicolumn{3}{c|}{\textbf{Clotho}} & \multicolumn{3}{c}{\textbf{MECAT}} \\
& R@1 & R@5 & R@10 & R@1 & R@5 & R@10 & R@1 & R@5 & R@10 \\
\midrule
LAION-CLAP & 28.51 & 62.56 & 76.92 & 17.51 & 41.63 & 55.02 & 7.31 & 23.94 & 33.49 \\
Robust-CLAP & 31.38 & 67.90 & 83.28 & 18.37 & 47.37 & 61.44 & 8.37 & 26.30 & 36.44 \\
MGA-CLAP & 41.23 & \uline{76.31} & 87.08 & 22.11 & 50.43 & 63.64 & 8.37 & 26.18 & 37.85 \\
M2D-CLAP & \textbf{46.46} & \textbf{82.77} & \textbf{93.64} & 22.11 & 48.52 & 62.49 & 8.84 & \textbf{28.18} & \textbf{40.21} \\
\midrule
OEA-Nemo3B & 40.51 & 73.44 & 85.44 & 21.53 & 45.45 & 59.14 & \uline{9.08} & \uline{27.71} & \uline{39.74} \\
OEA-Nemo3B (+Cl) & 36.92 & 69.64 & 83.28 & 25.93 & 52.54 & 66.03 & 7.90 & 24.88 & 37.03 \\
OEA-Qwen3B & 38.97 & 73.03 & 85.85 & 24.50 & 49.76 & 61.72 & 7.55 & 21.82 & 29.60 \\
OEA-Qwen3B (+Cl) & 39.08 & 72.51 & 84.51 & \uline{27.66} & \textbf{57.99} & \textbf{71.87} & 6.25 & 19.69 & 27.24 \\
OEA-Qwen7B & \uline{43.79} & 74.87 & \uline{88.00} & 24.98 & 50.72 & 65.93 & \textbf{9.79} & 25.35 & 36.56 \\
OEA-Qwen7B (+Cl) & 38.15 & 70.46 & 83.08 & \textbf{27.75} & \uline{55.41} & \uline{68.61} & 8.25 & 24.76 & 36.20 \\
\bottomrule
\end{tabular}
\caption{Keyphrase query T2A results (\%). \textbf{Bold}: best per column, \uline{underline}: second best.}
\label{tab:tagging_detailed}
\end{table*}


\section{Hard Negative Mining Pipeline}
\label{sec:hn_mining_appendix}

We develop a four-stage pipeline to construct hard negative pairs for exclusionary query evaluation:

\paragraph{Stage 1: Acoustic Candidate Retrieval}
For each target audio, we retrieve top-$K$ (K=20) acoustically similar candidates using MGA-CLAP audio embeddings.

\paragraph{Stage 2: Acoustic Similarity Filtering}
We apply a dynamic threshold $\theta_{\text{acoustic}}$ to retain candidates with high acoustic similarity, keeping approximately $3\times$ the final target count. This threshold is calibrated via human listening on sampled pairs (Appendix~\ref{sec:hn_human_validation}).

\paragraph{Stage 3: Semantic Dissimilarity Scoring}
We compute semantic similarity between target and candidate captions using BGE-large-en-v1.5 \cite{xiao2023bge} sentence embeddings.

\paragraph{Stage 4: Semantic Dissimilarity Filtering}
We retain pairs where captions are sufficiently different ($\sim$1$\times$ target count), yielding audio clips that are \textit{acoustically similar but semantically distinct}---ideal hard negatives for exclusionary queries. The dissimilarity cutoff is likewise calibrated on sampled pairs (Appendix~\ref{sec:hn_human_validation}).


\section{Additional Evaluation Metrics}
\label{sec:additional_metrics}

In addition to HNSR@k (reported in main text), we introduce the following metrics for exclusionary query evaluation:

\begin{itemize}
\item \textbf{$\Delta$-Rank}: $\text{Rank}(\text{HN}) - \text{Rank}(\text{T})$. Higher values indicate better separation between target (T) and hard negative (HN). This provides a continuous measure of separation quality suitable for model comparison.

\item \textbf{HNSR} (Hard Negative Suppression Rate): Percentage of queries where $\text{Rank}(\text{HN}) > \text{Rank}(\text{T})$---i.e., target ranked above hard negative, regardless of absolute rank.

\item \textbf{TFR} (Target-First Rate): Percentage where target is ranked first.

\item \textbf{TFR-HN@k}: Percentage where target is ranked first AND hard negative is outside top-$k$---the strictest combined metric for applications requiring high-confidence exclusion understanding.
\end{itemize}


\section{Negative Query Detailed Results}
\label{sec:negative_appendix}

Table~\ref{tab:negative_appendix} presents comprehensive negative query evaluation results including all discrimination metrics.

\begin{table*}[h]
\centering
\small
\begin{tabular}{l|cc|ccc|cc}
\toprule
\multirow{2}{*}{\textbf{Model}} & \multicolumn{2}{c|}{\textbf{Standard Retrieval}} & \multicolumn{3}{c|}{\textbf{Hard Negative Discrimination}} & \multicolumn{2}{c}{\textbf{Strict Metrics}} \\
& R@5 & R@10 & $\Delta$-Rank & HNSR & HNSR@10 & TFR & TFR@10 \\
\midrule
LAION-CLAP & 30.72 & 42.15 & 25.6 & 68.7 & 24.7 & 10.5 & 4.9 \\
Robust-CLAP & 34.86 & 46.53 & \textbf{33.7} & 69.6 & 30.3 & 11.6 & 6.9 \\
MGA-CLAP & 36.75 & 48.92 & 15.1 & 65.0 & 21.2 & 12.9 & 3.9 \\
M2D-CLAP & \textbf{41.56} & \textbf{55.63} & 18.6 & 68.0 & 30.3 & 16.2 & 7.5 \\
\midrule
OEA-Nemo3B & 39.64 & 51.23 & 21.1 & 68.6 & \uline{32.9} & 15.1 & 8.6 \\
OEA-Nemo3B (+Cl) & 41.48 & 53.67 & 17.9 & 69.2 & 32.5 & \textbf{17.3} & \textbf{10.1} \\
OEA-Qwen3B & 37.49 & 49.12 & 18.8 & 67.5 & 31.6 & 14.3 & 8.3 \\
OEA-Qwen3B (+Cl) & \uline{41.52} & \uline{53.89} & 17.6 & 67.3 & 32.8 & \uline{16.4} & \uline{10.0} \\
OEA-Qwen7B & 38.20 & 50.45 & \uline{33.0} & \textbf{74.4} & 31.8 & 15.8 & 9.5 \\
OEA-Qwen7B (+Cl) & 38.14 & 50.38 & 29.8 & \uline{73.9} & \textbf{34.6} & 15.9 & 9.8 \\
\bottomrule
\end{tabular}
\caption{Negative query evaluation (mean across datasets). Standard Retrieval: R@k (\%). Hard Negative Discrimination: $\Delta$-Rank (rank gap), HNSR/HNSR@10 (\%). Strict Metrics: TFR/TFR@10 (\%). M2D-CLAP achieves best standard retrieval, while OEA models substantially outperform all baselines on discrimination metrics. \textbf{Bold}: best, \uline{underline}: second best.}
\label{tab:negative_appendix}
\end{table*}


\section{Inference Efficiency Analysis}
\label{sec:efficiency_analysis}

A practical concern for deploying LLM-based retrieval models is computational cost relative to lightweight CLAP baselines. We conduct a systematic efficiency benchmark on the Clotho evaluation set (1,045 clips) using an NVIDIA A100-SXM4-80GB GPU. Table~\ref{tab:efficiency} reports per-clip audio encoding latency, per-query text encoding latency, peak GPU memory, and the number of trainable parameters.

\begin{table}[h]
\centering
\footnotesize
\setlength{\tabcolsep}{4pt}
\begin{tabular}{l|cc|cc}
\toprule
\textbf{Model} & \textbf{Audio} & \textbf{Text} & \textbf{Peak} & \textbf{Train.} \\
& (ms/clip) & (ms/q) & (GB) & (M) \\
\midrule
LAION-CLAP & 107.7 & 0.53 & $\sim$0.6 & 158 \\
MGA-CLAP & 31.7 & 0.72 & $\sim$0.6 & 148 \\
M2D-CLAP & 58.1 & 0.3 & 0.7 & 89 \\
\midrule
OEA-Nemo3B & 163.8 & 2.3 & 11.5 & 13.7 \\
OEA-Qwen3B & 539.3 & 2.60 & 11.6 & 16.2 \\
OEA-Qwen7B & 666.8 & 4.86 & 18.3 & 17.2 \\
\bottomrule
\end{tabular}
\caption{Inference efficiency comparison on Clotho (1,045 clips, A100-SXM4-80GB). Audio: per-clip encoding latency. Text: per-query encoding latency. Peak GPU: peak GPU memory during inference. Train. Params: number of trainable parameters during LoRA fine-tuning.}
\label{tab:efficiency}
\end{table}

Several observations warrant discussion:

\paragraph{Audio Encoding is Offline and Amortized}
In practical retrieval systems, audio in the database is encoded offline and indexed once. Per-query latency therefore depends only on text encoding, where OEA-Nemo3B achieves 2.3 ms/query---sufficient for real-time serving.

\paragraph{LoRA Enables Parameter-Efficient Adaptation}
Despite using 3B--7B parameter backbones, OEA fine-tunes only 13.7--17.2M parameters (0.29--0.36\% of total), demonstrating efficient adaptation that requires substantially fewer trainable parameters than CLAP models (89--158M).

\paragraph{Architectural Design Drives Latency Differences}
The latency gap between OEA-Nemo3B and OEA-Qwen3B (163.8 vs. 539.3 ms/clip) reflects architectural design rather than model size. Both share an identical audio encoder (\texttt{qwen2\_5\_omni\_audio\_encoder}, 32 layers, $d$=1280). OEA-Nemo3B loads only the embedding-optimized thinker with bidirectional attention, while OEA-Qwen3B loads the full generative model (thinker + TTS talker + token2wav vocoder), incurring additional GPU memory-bandwidth pressure even though TTS components are inactive during retrieval.

\paragraph{Quantization for Consumer-Grade Deployment}
With 4-bit quantization (e.g., GPTQ/AWQ), OEA-Nemo3B reduces from $\sim$6 GB (BFloat16) to $\sim$1.5 GB, enabling single-GPU deployment on consumer hardware such as an NVIDIA RTX 4090 (24 GB).


\section{UIQ Benchmark Bias Analysis}
\label{sec:uiq_bias_analysis}

Since the UIQ benchmark is generated by GPT-5.1, a potential concern is that OEA models might systematically benefit from shared linguistic patterns with the generator model. We address this through multiple lines of evidence.

\paragraph{Human Validation Confirms Departure from Caption Style}
Nine annotators rated UIQ naturalness at 4.15/5.0 (Table~\ref{tab:uiq_validation}) with strong inter-annotator agreement ($r=0.89$ for imperative queries), specifically confirming that queries reflect genuine search intent rather than caption paraphrases.

\paragraph{Vocabulary Grounding as Controlled Design}
We constrained UIQ generation to reuse words from original captions ($\pm$2 words) to prevent length-induced distribution shift---a deliberate methodological choice for controlled evaluation. The key variation tested is query \textit{formulation style} (questions, commands, keyphrases, paraphrases, negations), not vocabulary novelty.

\paragraph{Empirical Evidence Against Systematic Bias}
If OEA systematically benefited from GPT-generated linguistic patterns, it should dominate \textit{all} UIQ types. Instead, M2D-CLAP achieves the highest Avg UIQ (47.76\% vs. OEA-Qwen3B's 47.18\%, Table~\ref{tab:uiq_mean}), and OEA's advantages concentrate in Imperative queries and hard negative discrimination---tasks where instruction-following pretraining and the discrimination of OEA's audio-text embedding space provide a principled advantage (Appendix~\ref{sec:mechanism_ablations}).

\paragraph{Future Directions}
To further strengthen the claim that GPT-generated queries represent human intent, future work should compare UIQ token distributions against real-world audio search logs, such as Freesound user queries analyzed by \citet{weck2024language}, which average 1.8 tokens and exhibit substantially different distributional characteristics from caption-style text.


\section{Scaling Behavior and Multi-Backbone Generality}
\label{sec:scaling_behavior}

\subsection{Multi-Backbone Evaluation}

To demonstrate generality across architectures, we evaluate OEA with three distinct backbone LLMs from two organizations:

\begin{itemize}
    \item \textbf{Nemotron-3B} (NVIDIA): Embedding-optimized architecture with bidirectional attention
    \item \textbf{Qwen2.5-Omni-3B} (Alibaba): Full generative multimodal model at 3B scale
    \item \textbf{Qwen2.5-Omni-7B} (Alibaba): Larger-scale variant of Qwen2.5-Omni
\end{itemize}

These backbones span different pretraining corpora, architectural designs, and parameter scales. Consistent trends across all three backbones---including dominant T2T performance and superior hard negative discrimination relative to CLAP baselines---strengthen the generalizability of our findings. Native audio understanding is a hard architectural requirement for our unified encoder design; adopting text-only LLMs would necessitate a separate audio encoder, fundamentally changing the architecture. Extension to additional native-audio LLMs as they emerge remains an important direction for future work.

\subsection{Retrieval-Specific Scaling Behavior}

An interesting observation is that retrieval performance does not consistently improve with increased model scale. On MECAT T2A R@1, OEA-Nemo3B achieves 7.96\% compared to OEA-Qwen7B's 7.74\%, and on certain UIQ metrics the 3B models match or exceed 7B performance.

This pattern is consistent with prior findings in dense retrieval literature. Studies on embedding models such as E5 \cite{wang2024text} and BGE \cite{xiao2023bge} have reported diminishing returns above $\sim$3B parameters for embedding tasks, suggesting that the bottleneck shifts from backbone capacity to \textit{contrastive alignment quality} at larger scales. In our setting, the training data volume and contrastive learning dynamics appear to constrain how effectively larger backbones can leverage their additional parameters.

The 7B model's marginal underperformance on some metrics further suggests that \textit{dataset--backbone alignment} matters more than raw parameter count for retrieval tasks. This finding has practical implications: the 3B model achieves competitive or superior performance at substantially lower computational cost (Table~\ref{tab:efficiency}), making it the recommended configuration for deployment.


\section{Hard Negative Human Validation}
\label{sec:hn_human_validation}

A potential limitation of our hard negative mining pipeline is its reliance on MGA-CLAP for acoustic similarity retrieval (Stage 1), which constrains the definition of ``hard'' to the representational capacity of MGA-CLAP. If MGA-CLAP fails to identify certain types of acoustic similarity, those confusion types would be omitted from the evaluation set.

To mitigate this single-model dependency, we implemented two safeguards:

\paragraph{Dual-Model Filtering}
The pipeline combines MGA-CLAP (acoustic similarity) with BGE-large-en-v1.5 \cite{xiao2023bge} (semantic dissimilarity), reducing reliance on any single model's representations. This ensures that selected pairs are both acoustically similar (per MGA-CLAP) and semantically distinct (per BGE), providing complementary perspectives.

\paragraph{Human-Guided Threshold Calibration}
Human judgment enters the pipeline through threshold calibration rather than exhaustive per-pair auditing. During pipeline development, annotators listened to \textit{sampled} target--candidate pairs produced under candidate settings of the acoustic similarity threshold ($\theta_{\text{acoustic}}$, Stage~2) and the semantic dissimilarity cutoff (Stage~4). The thresholds were adjusted until sampled pairs were consistently judged perceptually confusable---acoustically similar to human ears yet semantically distinct---and the final thresholds were then applied uniformly to mine the full set. We did not manually audit every mined pair; the human signal shapes the selection criteria rather than the individual selections. This calibration reduces the risk that the benchmark merely rewards agreement with MGA-CLAP's embedding space, while remaining scalable to the full benchmark size.

While these measures reduce the impact of MGA-CLAP's representational limitations, we acknowledge that some perceptually confusing pairs may still be missed if they fall entirely outside MGA-CLAP's embedding space. Future work could explore ensemble retrieval using multiple audio encoders with diverse architectures to broaden the coverage of acoustic confusion types.


\section{Mechanism Ablations for Hard Negative Discrimination}
\label{sec:mechanism_ablations}

The main text attributes OEA's hard negative discrimination advantage to its LLM backbone. To decompose this claim, we run three post-hoc ablations on the released OEA-Qwen7B (+Cl) checkpoint with \textbf{no retraining}, comparing against M2D-CLAP (cleanest baseline), MGA-CLAP (the hard-negative \textit{miner}), and LAION-CLAP on the paper's mined (target, hard negative) pairs across AudioCaps (630 pairs), Clotho (542), and MECAT (409). We separate an \textbf{audio path} (are target T and hard negative HN separable in audio embedding space at all?) from a \textbf{text path} (how does the backbone process the exclusion clause?).

\subsection{Ablation 1: Audio-Only Separation}
\label{sec:ablation_audio}

Independent of any query, we measure the audio-embedding cosine distance $d = 1 - \cos(\mathbf{e}_T, \mathbf{e}_{HN})$ and the corpus rank of HN when the gallery is sorted by similarity to T (higher = the audio space already pushes HN away).

\begin{table}[h]
\centering
\small
\begin{tabular}{l|ccc|c}
\toprule
\textbf{Model} & \textbf{AC} & \textbf{Cl} & \textbf{ME} & \textbf{Mean} \\
\midrule
\multicolumn{5}{c}{\textit{T--HN audio distance $d$ (higher = better)}} \\
\midrule
OEA-Qwen7B (+Cl) & \textbf{0.334} & \textbf{0.401} & \textbf{0.215} & \textbf{0.317} \\
M2D-CLAP & 0.273 & 0.215 & 0.180 & 0.223 \\
MGA-CLAP (miner) & 0.331 & 0.284 & 0.192 & 0.269 \\
LAION-CLAP & 0.358 & 0.356 & 0.219 & 0.311 \\
\midrule
\multicolumn{5}{c}{\textit{Corpus HN-rank (higher = HN pushed farther)}} \\
\midrule
OEA-Qwen7B (+Cl) & \textbf{27.2} & \textbf{53.8} & \textbf{28.7} & \textbf{36.6} \\
M2D-CLAP & 21.6 & 32.9 & 24.8 & 26.4 \\
MGA-CLAP (miner) & 9.5 & 16.7 & 11.7 & 12.6 \\
LAION-CLAP & 18.8 & 27.2 & 24.9 & 23.7 \\
\bottomrule
\end{tabular}
\caption{Ablation 1: audio-only separation of mined (T, HN) pairs. AC: AudioCaps, Cl: Clotho, ME: MECAT. Paired Wilcoxon signed-rank tests on the same pairs confirm OEA $>$ M2D-CLAP on every dataset: $\Delta d = +0.061$ ($p = 1.6\times10^{-22}$), $+0.187$ ($p = 1.7\times10^{-68}$), $+0.035$ ($p = 1.1\times10^{-7}$) on AC/Cl/ME respectively.}
\label{tab:ablation1}
\end{table}

OEA separates confusable clips significantly better than M2D-CLAP on every dataset and attains the highest corpus HN-rank of any model. MGA-CLAP's small distances are the expected miner confound---it defined the pairs on its own acoustic similarity---which is precisely why OEA-vs-M2D is the intended clean comparison. \textbf{The audio path genuinely contributes to discrimination.}

\subsection{Ablation 2: Negation On/Off}
\label{sec:ablation_negation}

We run each exclusionary query as-is (NEG, e.g., ``male speech \textit{without} a female voice'') and with the exclusion clause stripped at the negation cue (POS, ``male speech''); the positive content is byte-identical. $\Delta\Delta$-Rank $=$ $\Delta$-Rank(NEG) $-$ $\Delta$-Rank(POS) measures the net effect of the clause.

\begin{table}[h]
\centering
\small
\begin{tabular}{l|cc|c}
\toprule
\textbf{Model} & \textbf{NEG} & \textbf{POS} & \textbf{$\Delta\Delta$-Rank} \\
\midrule
OEA-Qwen7B (+Cl) & 22.5 & 55.2 & $-32.7$ \\
M2D-CLAP & 17.0 & 52.8 & $-35.8$ \\
MGA-CLAP (miner) & 15.9 & 42.6 & $-26.8$ \\
LAION-CLAP & 31.1 & 50.6 & $-19.5$ \\
\bottomrule
\end{tabular}
\caption{Ablation 2: $\Delta$-Rank with (NEG) and without (POS) the exclusion clause, mean across datasets. The clause \textit{hurts every model}: it lexically mentions the excluded content, and the resulting pull toward the HN is not fully cancelled by the negation token---not even for OEA. OEA is, however, the most robust on the strict metric ($\Delta$HNSR@10: $-6.9$ vs.\ M2D's $-14.9$ and MGA's $-11.6$) and posts the best HNSR@10 on the actual exclusionary task (34.4).}
\label{tab:ablation2}
\end{table}

This result sharpens the mechanism claim: OEA does \textbf{not} convert the negation clause into a clean downward push on the excluded clip. Its advantage on exclusionary queries is robustness to the clause's lexical pull, not sign-flipping of the excluded term.

\subsection{Ablation 3: Compositional vs.\ Bag-of-Content Text Processing}
\label{sec:ablation_boc}

Keeping OEA's audio branch and projection heads fixed, we shuffle query word order before the transformer. The shuffle preserves the exact word multiset (verified on all queries; ``without'' remains present) and destroys \textit{only} order/structure. If compositional structure carries the exclusion signal, shuffling should hurt exclusionary queries specifically.

\begin{table}[h]
\centering
\small
\begin{tabular}{l|cc|c}
\toprule
\textbf{Query content} & \textbf{Full} & \textbf{Shuffled} & \textbf{Gap} \\
\midrule
Exclusionary (NEG) & 22.5 & $-7.2$ & $+29.7$ \\
Positive-only (POS) & 55.2 & 60.5 & $-4.6$ \\
\bottomrule
\end{tabular}
\caption{Ablation 3: OEA-Qwen7B (+Cl) $\Delta$-Rank with full contextual encoding vs.\ word-shuffled queries (mean across datasets; gap $=$ full $-$ shuffled). Destroying word order collapses exclusionary separation---flipping the mean HN margin negative---while leaving positive queries essentially unchanged (within noise). HNSR@10 gaps show the same pattern ($+18.4$ exclusionary vs.\ $+4.8$ positive).}
\label{tab:ablation3}
\end{table}

Because the tokens are identical and only their order changes, this isolates structure unambiguously: \textbf{attention over query structure---not the negation word alone---carries the exclusion signal} within OEA. A cruder pre-transformer mean-pooling control destroys retrieval entirely (R@1 $\approx 0$) and is reported only as a floor. Ablation~3 alone, however, cannot establish that this reliance on structure is unique to OEA; we test that next.

\subsection{Ablation 3-X: Cross-Model Shuffle Control}
\label{sec:ablation_boc_cross}

We apply the byte-identical, per-query pinned shuffle to the CLAP baselines and compute each model's exclusion-specific structure contribution $S(m) = \text{gap}_{\text{excl}}(m) - \text{gap}_{\text{pos}}(m)$, where gap $=$ $\Delta$-Rank(full) $-$ $\Delta$-Rank(shuffled). All models see the same shuffled text and the same pairs, so per-query paired tests are valid.

\begin{table}[h]
\centering
\small
\setlength{\tabcolsep}{4pt}
\begin{tabular}{l|cc|c|c}
\toprule
\textbf{Model} & \textbf{gap$_{\text{excl}}$} & \textbf{gap$_{\text{pos}}$} & \textbf{$S(m)$} & \textbf{rel.} \\
\midrule
LAION-CLAP & 35.9 & $-13.2$ & \textbf{49.0} & 1.15 \\
OEA-Qwen7B (+Cl) & 29.7 & $-4.6$ & \textbf{34.3} & 1.32 \\
M2D-CLAP & 23.8 & $-7.8$ & \textbf{31.6} & 1.40 \\
MGA-CLAP (miner) & 21.7 & $-4.0$ & \textbf{25.8} & 1.37 \\
\bottomrule
\end{tabular}
\caption{Ablation 3-X: exclusion-specific structure contribution $S(m)$, mean across datasets; rel.\ $=$ gap$_{\text{excl}}$ normalized by the full-order exclusionary margin. Every model's exclusionary shuffle damage is highly significant (Wilcoxon $p \ll 10^{-3}$ on every dataset). Paired per-query comparisons of $S$: OEA vs.\ M2D $+4.1$ ($p=0.017$), vs.\ MGA $+9.6$ ($p=1.9\times10^{-4}$), vs.\ LAION $-17.3$ ($p=5\times10^{-4}$).}
\label{tab:ablation3x}
\end{table}

Order sensitivity is a \textbf{general property of transformer text encoders}, not an OEA-specific mechanism: every encoder collapses on exclusionary queries when word order is destroyed. OEA's structure contribution (34.3) exceeds M2D's (31.6, weakly significant) and MGA's (25.8) but is smaller than LAION's (49.0), and in relative terms OEA's collapse (1.32) is comparable to or below the CLAP baselines (1.40 / 1.37). The text side of OEA is therefore on par with strong CLAP encoders; the decisive, OEA-specific advantage remains the audio-side separation of Ablation~1.

\subsection{Summary}

OEA's hard negative advantage rests on a well-formed \textbf{audio space} (Ablation~1: OEA $>$ M2D on every dataset, all $p \ll 0.001$) combined with an order-sensitive text encoding that uses query structure for exclusion (Ablation~3). It is \textit{not} a lexical ``negation word $\to$ downward push'' mechanism (Ablation~2: the exclusion clause hurts every model, OEA included), and it is \textit{not} a text-side mechanism CLAP lacks (Ablation~3-X: order sensitivity is general, and OEA's structure contribution sits between M2D/MGA and LAION). OEA's distinctive edge is the \textbf{magnitude of separation} it achieves---driven chiefly by its audio embeddings---together with the best strict-metric performance on the actual exclusionary task (HNSR@10). This refines the hypothesis stated in earlier versions of this paper, which attributed the advantage to exposure to negation patterns during instruction-following pretraining.

\end{document}